\documentclass[lettersize,journal]{IEEEtran}
\usepackage{mathtools}
\usepackage{amssymb}
\usepackage{enumerate}
\usepackage{definice}
\usepackage{color}
\usepackage{diagbox,multirow}
\usepackage{multirow}
\usepackage{float}
\usepackage{comment}
\usepackage{cite}

\usepackage{amsmath,amsfonts}
\usepackage{algorithmic}
\usepackage{algorithm}
\usepackage{array}
\usepackage[caption=false,font=normalsize,labelfont=sf,textfont=sf]{subfig}
\usepackage{textcomp}
\usepackage{stfloats}
\usepackage{url}
\usepackage{verbatim}
\usepackage{graphicx}

\usepackage{enumitem}

\usepackage[para]{threeparttable}
\usepackage{booktabs}

\usepackage{url}
\usepackage{hyperref}
\usepackage{xspace}
\newcommand{\MATLAB}{\textsc{Matlab}\xspace}

\usepackage[scr=boondoxo,scrscaled=1.05]{mathalfa}

\usepackage{mdframed}    
\newenvironment{infobox}[1]
{
	\mdfsetup{
		frametitle={\colorbox{white}{\space#1\space}},
		frametitleaboveskip=-\ht\strutbox,
	}
	\begin{mdframed}
	}{
	\end{mdframed}
}

\usepackage{amsfonts} 
\DeclareMathSymbol{\shortminus}{\mathbin}{AMSa}{"39}

\def\p{\hspace{1mm}\mathbf{\grave\!} L}

\def\zkl{z\hspace{-0.2mm},\hspace{-0.2mm}k\hspace{-0.2mm},\hspace{-0.2mm}L}
\def\uklm{u\hspace{-0.2mm},\hspace{-0.2mm}k\hspace{-0.2mm},\hspace{-0.2mm}\p}

\def\rmU{\mathrm U}

\def\aO{n_{\mathsf{a}\raisebox{-0.45mm}{\scriptsize$\calO$}}}

\def\aOG{n_{\mathsf{a}\raisebox{-0.45mm}{\scriptsize$\calO\calG$}}}

\def\re{n_\calE}

\usepackage{graphicx}      

\newcommand{\tightoverset}[2]{%
  \mathop{#2}\limits^{\vbox to -.5ex{\kern-0.75ex\hbox{$#1$}\vss}}}

\begin{document}
\title{Unobservable Systems: No Problem\\ for Noise Identification}
\author{Oliver Kost, Jind\v{r}ich Dun\'{i}k, Ivo Pun\v{c}ocha\v{r}, Ond\v{r}ej Straka
\thanks{Authors are with Department of Cybernetics, Faculty of Applied Sciences, University of West Bohemia in Pilsen, Univerzitn\'{i} 8, 306 14 Pilsen, Czech Republic. E-mails:  \{kost,dunikj,ivop,straka30\}@kky.zcu.cz.
The work has been partially supported by the Ministry of Education, Youth and Sports of the Czech Republic under project ROBOPROX - Robotics and Advanced Industrial Production CZ.02.01.01/00/22\_008/0004590.
}
}
\maketitle
\begin{abstract} 
This paper deals with the noise identification of a linear time-varying stochastic dynamic system described by the state-space model. In particular, the stress is laid on the design of the correlation \textit{measurement difference method} for estimation of the state and measurement noise covariance matrices for both \textit{observable} and \textit{unobservable} systems with possibly unknown input sequence. The method provides unbiased and consistent estimates and is implemented in a publicly available \MATLAB toolbox and numerically evaluated.
\end{abstract}


\section{Introduction}

The availability of an accurate system model is a key prerequisite for the design of reliable algorithms for optimal control, fault detection, and signal-processing algorithms. Negligent system modelling may lead to insufficient control, detection, estimation quality, or even algorithm failure.

Modern algorithms rely on the availability of the stochastic state-space model \cite{GaSaClGo:22}. The model describes deterministic and stochastic phenomena governing the behaviour of the system quantities (called state) and the relation between the unknown state and the available noisy measurement produced by sensors. Hence, the state-space model can be virtually seen as a composition of two sub-models: \textit{deterministic} and \textit{stochastic}. The former captures system behaviour arising often from the first principles, based on physical, kinematical, chemical, mathematical, biological, or other laws and rules. The latter characterises the uncertain behaviour of the state and measurement, where the uncertainty is modelled by noises described in the stochastic framework. Due to the inherent difficulty in uncertainty modelling using the first principles, it has to be identified from data.

Therefore, since the seventies, intensive research has been laid on the design of methods identifying properties, typically the covariance matrices, of the state and measurement noises. In the literature, five classes of noise identification methods are recognised \cite{Meh:72,DuStKoHa:17}, namely, Bayesian \cite{SaNu:09,HuZyShCh:21}, covariance matching \cite{MoKi:93,SoHaIlAbBi:14}, \textcolor{black}{maximum likelihood \cite{ShSt:00,ScWiNi:11,LuLePu:21}}, machine-learning-based \cite{CoKl:24}, and correlation methods.

In this paper, we focus on the correlation methods \cite{Be:74,OdeRaRa:06,KoDuSt:22}, which, compared with other classes, are derived analytically without any assumption on the noise distribution or structure. These methods rely on the assumption of the system \textit{observability} or \textit{detectability} and \textit{known system input} (if the input is present), which might be limiting for certain applications. For example, the \textit{unobservable} system can be found in generation of the stable time scale\footnote{The accurate time synchronisation is required in a wide range of applications ranging from navigation through power distribution to financial services. The time scale is generated from an ensemble of (typically) atomic clocks using the Kalman-filter-based combination algorithms. As such, the accurate (linear) stochastic state-space model of the clock ensemble is required, where the deterministic part stems from the underlying physics and the stochastic part is identified from data \cite{Galleani2008,Galleani2010}.}, where the lack of observability is caused by measuring clock phase differences only \cite{Galleani2010}. The system with \textit{unknown input} occurs, e.g., in a tracking, where the tracked object thrust \textit{need not} be available \cite{BaLiKi:01}. 



In particular, we \textit{propose} ordinary and weighted generalisations of the correlation \textit{measurement difference method} (MDM) \cite{KoDuSt:22} allowing noise identification of linear stochastic system with \textit{unobservable} state and \textit{unknown} input sequence. Both generalisations, which differ in the estimation accuracy and algorithm complexity, adopt parametrisation of the unknown noise covariance matrices using a set of structure-defining matrices from \cite{Be:74}. These matrices allow simple and efficient descriptions of almost any noise structure. The developed MDM is implemented in available \MATLAB codes.

\section{System Description and Noise Identification}  \label{sec:noise_covariance_identification}
Consider the following linear time-varying (LTV) discrete-in-time stochastic dynamic model in state-space representation with additive noises and varying measurement dimension \!\!\!\!\cite{LiDeCh:19}\!\!\!
\begin{subequations}\label{model}
\begin{align}
\bfx_{k+1}&=\bfF_k\bfx_k+\bfG_k\bfu_k+\bfE_k\bfw_k,\label{eqSt}\\
\bfz_{k}&=\bfH_k\bfx_k+\bfD_k\bfv_k,\label{eqMe}
\end{align}
\end{subequations}
for $ k=0,1,2,\ldots,\tau $. The vectors $\bfx_k\in\real^{n_x}$, $\bfu_k\in\real^{n_{u_k}}$, and $\bfz_k\in\real^{n_{z_k}}$ represent the immeasurable state of the system, the input, and the available measurement at time instant $k$, respectively. The state $\bfF_k\in\real^{n_x\times n_x}$, input $\bfG_k\in\real^{n_x\times n_{u_k}}$, state-noise-shaping $\bfE_k\in\real^{n_x\times n_w}$, measurement $\bfH_k\in\real^{n_{z_k}\times n_x}$, and measurement-noise-shaping $\bfD_k\in\real^{n_{z_k}\times n_v}$ matrices are known and bounded. 
The state noise $\bfw_k\in\real^{n_w}$ and the measurement noise $\bfv_k\in\real^{n_v}$ are assumed to be zero-mean random variables with \textit{unknown} but bounded covariance matrices $\bfQ\in\real^{n_w\times n_w}$ and $\bfR\in\real^{n_v\times n_v}$.




\subsection{Noise Identification}
The noise identification aims at the estimation of the covariance matrices $\bfQ$ and $\bfR$ from a set of measurements $\bfz_k$ (and inputs $\bfu_k$ if available) and known matrices $\bfF_k, \bfG_k, \bfE_k, \bfH_k, \bfD_k, \forall k$.
\subsection{Measurement Difference Method and the Goal}
The \textit{measurement difference method} is a correlation method for the noise properties identification \cite{DuKoSt:18,KoDuSt:22}. 
It is based on a statistical analysis of the available \textit{residual} (or measurement prediction error) vector designed to be a \textit{linear} function of the unknown state and measurement noise vectors. Availability of the residual vector allows noise identification (e.g., the noise covariance matrix (CM) estimation) using the \textit{least-squares} (LS) method. Contrary to other correlation methods, the MDM provides unbiased and consistent estimates of the noise moments for an LTV model.

The MDM, as well as any state-of-the-art noise identification method, relies on the assumptions of observable state $\bfx_k$ and known input $\bfu_k, \forall k$ often under the assumption of time-invariant dimensions of the measurement and the input. 
%
In this paper, we generalise the MDM by \textit{relaxing} the limiting assumptions on state observability, input availability, and invariant dimensions. 

In the following sections, we 
\begin{itemize}[leftmargin=5mm]
    \item Develop the residue calculation and analyse its statistical properties w.r.t. the state and measurement noises (Sec.~\ref{sec:residue}).
    \item Adopt noise CM parametrisation using a sum of known structure-defining matrices with unknown (and sought) weights that allow the efficient description of a priori known complex noise structures (Sec.~\ref{sec:residue}).
    \item Develop two MDM versions for noise covariance estimation differing in accuracy and complexity (Sec.~\ref{sec:mdm_identification_techniques}).
    \item Analyse the MDM, relate it to other correlation methods and extend the MDM for identification under the assumption of an unknown input vector (Sec.~\ref{sec:mdm_analysis}).    
\end{itemize}

\section{Residue Definition, Calculation, and its Covariance}\label{sec:residue}
The unique properties of the MDM stem from the definition of the residue, which can be calculated from available measurements and shown to be a linear function of the immeasurable state and measurement noises.



\subsection{Augmented Measurement Vector}
Derivation of the MDM starts with a definition of the augmented measurement vector as \cite{KoDuSt:22}
\begin{align}
	\bfZ_k = \calO_{k}\bfx_k + \bfGamma_k\calG_k\bfU_k + \bfGamma_k\scrE_k\bfW_k + \calD_k\bfV_k \label{Z}
\end{align}
for $k=0,\ldots,\tau-\p$, where $\p\!=\!L-1$ with $L\!\geq\!1$ {is a user-defined parameter of which selection is discussed later}, $\bfZ_k\in\real^{n_{\zkl}}$, $\calO_k\in\real^{n_{\zkl}\times n_x}$, $\bfGamma_k\in\real^{n_{\zkl}\times \p n_x}$, $\calG_k\in\real^{\p n_x\times n_{\uklm}}$, $\scrE_k\in\real^{\p n_x\times \p n_w}$, $\calD_k\in\real^{n_{\zkl}\times Ln_v }$, $\bfU_k\in\real^{n_{\uklm}}$, $\bfW_k\in\real^{\p n_w}$, and $\bfV_k\in\real^{Ln_v}$ are defined 
as 
\begin{subequations}\label{eq:matrices}
\begin{align}
	\!\!\!\!\bfZ_k&\triangleq\left[\begin{smallmatrix}\bfz_k \\ \bfz_{k+1} \\ \bfz_{k+2} \\[-2px] \vdots \\[-2px] \bfz_{k+\p }\end{smallmatrix}\right]\!,\ 
	\calO_k\triangleq\left[\begin{smallmatrix}
		\bfH_k \\ \bfH_{k+1}\bfF_{k} \\ \bfH_{k+2}\bfF_{k+1}\bfF_{k} \\[-3px] \vdots \\[-0px] \bfH_{k+\p }\calF_{k}^{\p}
		\hspace{-2px}
	\end{smallmatrix}\right]\!,\\
    \!\!\!\!\calG_k&\triangleq\blkdiag\begin{bmatrix}
		\bfG_{k}, & \bfG_{k+1}, \bfG_{k+2}, & \cdots &, \bfG_{k+L-2} \end{bmatrix}\!,
    \\
    \!\!\!\!\scrE_k&\triangleq\blkdiag\begin{bmatrix}
		\bfE_{k}, & \bfE_{k+1}, & \bfE_{k+2}, & \cdots &, \bfE_{k+L-2} \end{bmatrix},
    \\
    \!\!\!\!\calD_k&\triangleq\blkdiag\begin{bmatrix}
		\bfD_{k}, & \bfD_{k+1}, & \bfD_{k+2}, & \cdots &, \bfD_{k+\p} \end{bmatrix},
    \end{align}
    \begin{align}
	\!\!\!\!\bfGamma_k&\triangleq\left[\begin{smallmatrix}\bfnul_{n_{z_k}\times n_x} & \bfnul_{n_{z_k}\times n_x}& \cdots & \bfnul_{n_{z_k}\times n_x} 
		\\ \bfH_{k+1} & \bfnul_{n_{z_{k+1}}\times n_x} & \cdots & \bfnul_{n_{z_{k+1}}\times n_x} 
		\\ \bfH_{k+2}\bfF_{k+1} & \bfH_{k+2} & \cdots & \bfnul_{n_{z_{k+2}}\times n_x}
		\\[-2px] \vdots & \vdots & \hspace{-3px}\ddots & \vdots
		\\[-0px] \bfH_{k+\p }\calF_{k+1}^{L-2} & \bfH_{k+\p }\calF_{k+2}^{L-3} & \hspace{-0px}\cdots\hspace{-0px}& \bfH_{k+\p } \end{smallmatrix}\right]\!,\!\!\!
    \\
    \!\!\!\!\bfU_k&\triangleq\!\left[\begin{smallmatrix}\bfu_k \\ \bfu_{k+1} \\ \bfu_{k+2} \\[-3px] \vdots \\[-3px] \bfu_{k+L-2 }\end{smallmatrix}\right]\!,\ 
    \bfW_k\triangleq\!\left[\begin{smallmatrix}\bfw_k \\ \bfw_{k+1} \\ \bfw_{k+2} \\[-3px] \vdots \\[-3px] \bfw_{k+L-2 }\end{smallmatrix}\right]\!,\
    \bfV_k\triangleq\!\left[\begin{smallmatrix}\bfv_k \\ \bfv_{k+1} \\ \bfv_{k+2} \\[-3px] \vdots \\[-3px] \bfv_{k+\p}\end{smallmatrix}\right]\!,\!\!\!
\end{align}
\end{subequations}
where $\calF_{k}^{j}\triangleq\prod_{i=1}^j\hspace{-2px}\bfF_{k+j-i}=\bfF_{k+j-1}\hdots\bfF_{k+1}\bfF_{k}\in\real^{n_x\times n_x}$, $ n_{\zkl} = \Sigma_{i=0}^{\p }n_{z_{k+i}}$, $ n_{\uklm} = \Sigma_{i=0}^{L-2}n_{u_{k+i}}$, $\bfnul_{n_i\times n_j}$ denotes a zero matrix of dimension $n_i\times n_j$, and $\blkdiag[\cdot]$ denotes a block-diagonal matrix. 

\subsection{Residue Definition: Known Input}
The known augmented measurement vector $\bfZ_k$ \eqref{Z} depends on the unknown state, known input, and unknown state and measurement noises whose CM are sought. 
To eliminate the state, define the non-zero annihilation matrix $\anni(\calO_{k})\in\real^{\aO\times n_{\zkl}}$ of the matrix $\calO_{k}$ such that 
\begin{align}
    \anni(\calO_{k})\calO_{k}=\bfnul_{\aO\times n_x}.\label{eq:am}
\end{align}
where $\aO\!=\!{n_{\zkl}\!-\!\rank(\calO_{k})}$ and $\rank(\calO_{k})$ denotes a rank of the matrix $\calO_{k}$.
Then, the residue $\tbfZ_k\!\in\!\real^{\aO}$ is defined as\linebreak\\[-7mm]
\begin{subequations}
\begin{align}
	\tbfZ_k
	&= \anni(\calO_{k}) \left(\bfZ_k-\bfGamma_k\calG_k\bfU_k\right), \label{Zmdmnull}
	\\
	&=\anni(\calO_{k})\left(\bfGamma_k\scrE_k\bfW_k+\calD_k\bfV_k\right), \label{ZmdmnullWV}
\end{align}\label{Zmdmnullboth}
\end{subequations}
which further reads
\begin{subequations}
\label{mpe_all}
\begin{align}
	\!\!\!\tbfZ_k&=\calA_{k}\calB_{k}\left[\begin{smallmatrix}-\bfU_k\\\hspace{3mm}\bfZ_k\end{smallmatrix}\right]\!,\!\!\label{mpeMDMZU}
	\\
	\!&=\calA_{k}\calC_{k}\left[\begin{smallmatrix}\bfW_k \\\bfV_k \end{smallmatrix}\right]\!,\label{mpeMDMWV}
\end{align}
\end{subequations}
where the matrices $\calA_{k}\in\real^{\aO\times (\p n_x+n_{\zkl})}$,\linebreak $ \calB_{k}\in\real^{(\p n_x + n_{\zkl})\times (n_{\uklm} + n_{\zkl} )}$, $ \calC_{k}\in\real^{(\p n_x + n_{\zkl} )\times \re}$ and the vector $\calE_{k}\!\in\!\real^{\re}$ with $\re\!=\!\p n_w\!+\!Ln_v$ are defined as
\begin{subequations}
\begin{align} 
\!\!\calA_{k}&\triangleq \anni(\calO_{k})\begin{bmatrix}\bfGamma_k,&\!\! \bfI_{n_{\zkl}}\end{bmatrix}\label{eq:Awv}\!,\hspace{1.7mm} \calE_{k}\triangleq\begin{bsmallmatrix}\bfW_k \\ \bfV_k \end{bsmallmatrix}\!,\!\!
\\
\!\!\calB_{k} &\triangleq \blkdiag\begin{bmatrix}\calG_k,&\!\! \bfI_{n_{\zkl}}\end{bmatrix}\!,
\hspace{3mm}
\calC_{k} \triangleq \blkdiag\begin{bmatrix}\scrE_k,&\!\! \calD_k \end{bmatrix}\!.\!\!
\end{align}
\end{subequations}
where $\bfI_{n}$ denotes the identity matrix of dimension $n\times n$.
The residue $\tbfZ_k$ \eqref{mpeMDMZU} 
can be expressed explicitly as a weighted difference of the augmented \textit{measurement} vector and its prediction \cite{KoDuSt:22} or as a weighted difference of the augmented \textit{measurement} vector and augmented input vector as in \eqref{mpeMDMZU}.


\subsection{Relation of Residue and Noises Covariances}
Such definition of the residue $\tbfZ_k$ \eqref{mpe_all} leads to the following key properties. The residue is
\begin{itemize}
    \item A \textit{linear} function of the unavailable state and measurement noises as was shown in \eqref{mpeMDMWV},
    \item Calculated from the measurements \eqref{mpeMDMZU} even for \textit{unobservable} model, i.e., for a model with rank deficient observation matrix $\calO_k$.
\end{itemize}


Consequently, the \textit{time-varying} covariance of the residue for the time instant $k$ is a linear function of the \textit{time-invariant} CMs of the state and measurement noises, i.e.,
\begin{align}
	\calR_{\tbfZ_k^2}&\triangleq\mean\!\left[\tbfZ_k^{\otimes^2}\right]=
	\calA_{k}^{\otimes^2}\calC_{k}^{\otimes^2}
	\calR_{\calE^2},\label{noncentMom}
\end{align}
where $\otimes$ denotes the Kronecker product, $\calA^{\otimes^n}\!\!\!\triangleq\!\overbrace{\calA\!\otimes\!\ldots\!\otimes\!\calA}^{n-\text{terms}}$ denotes the Kronecker power, $\calR_{\tbfZ_k^2}\in\real^{\aO^2}$ is the vectorised residual \textit{covariance} containing the CM elements of the residue $\tbfZ_k$ and $\calR_{\calE^2}\in\real^{\re^2}$ contains elements of the CM of the augmented state and measurement noise vector $ \calE_{k} $.



\subsection{Noise CM Parametrisation via Structure-Defining Matrices}

The noise covariance $\calR_{\calE^2}$ contains multiple copies of \textit{unique} elements of the noise CMs $\bfQ$ and $\bfR$, due to either the construction of the residue, application of the Kronecker algebra, or the CM symmetry. For efficient estimation of the noise CM, it is necessary to express the \textit{unique} elements of the noise CMs $\bfQ, \bfR$ from the vector $\calR_{\calE^2}$ in \eqref{noncentMom}. An appealing parametrisation, introduced in \cite{Be:74}, defines $\bfQ, \bfR$ using a set of known (or user-defined) structure-defining matrices\footnote{{The structure-defining matrices are not unique and are task-specific. Their construction is illustrated in Section \ref{eq:exam1} for clock ensemble modelling \cite{Galleani2010}.}} $ \bfB_Q^{(i)}$ and $ \bfB_R^{(i)} $ and unknown (and sought) weights $\alpha_i$ as
\begin{align}
	\bfQ = \sum_{i=1}^{n_{\alpha}}\alpha_i\bfB_Q^{(i)},  \bfR = \sum_{i=1}^{n_{\alpha}}\alpha_i\bfB_R^{(i)}. \label{basMat}
\end{align}
Then, the noise CMs parametrisation allows to incorporate a priori knowledge of the noise structure using the \textit{minimum} number of noise CM parameters gathered in the vector denoted as $\bfalpha=[\alpha_1,\ldots,\alpha_{n_\alpha}]^T\in\real^{n_\alpha}$. The vectors $\calR_{\calE^2}$ and $\bfalpha$ are related via the defining-replication\footnote{If the matrices $\bfB_{Q}^{(i)} $ and $ \bfB_{R}^{(i)} $ contain only zeros and ones so that only unique elements of noise CMs are selected (as \eqref{basisR}), then the defining-replication matrix $\Upsilon_{\calE^2}$ is equivalent to the replication matrix $\Psi_{\calE^2}$ used in the articles on the MDM e.g., \cite{KoDuSt:22}.} matrix $\Upsilon_{\calE^2}\!\in\real^{\re^2\times n_{\alpha}}$ as
\begin{align}
	\calR_{\calE^2}&=\Upsilon_{\calE^2}\bfalpha,\label{NpsiM}
\end{align}
which, w.r.t. \eqref{basMat}, further reads 
\begin{align}
	\!\!\!\!\calR_{\calE^2}
	&\!=\!
    \left(\blkdiag\left[\bfI_{\p }\!\otimes\!\bfQ, \ \bfI_{L}\!\otimes\!\bfR\right]\right)_\mathsf{V} 
	= 
	\left(\mean\!\left[\calE_k \calE_k\T\right]\right)_\mathsf{V}
	\\
	&\!=\!
	\underbrace{\!
        \Bigg[\!
		\begin{smallmatrix}         
            \begin{bsmallmatrix}
				\!\bfI_{\p }\otimes\bfB_Q^{(1)} & \bfnul\\
				\bfnul & \bfI_{L}\otimes\bfB_R^{(1)}\!\!
			\end{bsmallmatrix}_\mathsf{\!V},
			&\!\!
			\begin{bsmallmatrix}
				\!\bfI_{\p }\otimes\bfB_Q^{(2)} & \bfnul\\
				\bfnul & \bfI_{L}\otimes\bfB_R^{(2)}\!\!
			\end{bsmallmatrix}_\mathsf{\!V},
            &  \ldots
		\end{smallmatrix}\!\Bigg]
	\!}_{\Upsilon_{\calE^2}}
	\hspace{1px}
	\underbrace{\!\!
		\begin{bsmallmatrix}
			\\[-0mm]
			\alpha_1
			\\
			\alpha_2
			\\[-1mm]
			\vdots
			\\[2mm]
		\end{bsmallmatrix}
		\!\!}_{\bfalpha}\!\ ,\!\!\!
\end{align}
where the notation $(\cdot)_\mathsf{V}$ stands for the column-wise stacking (or vectorisation) of the matrix and its inverse function is denoted as $(\cdot)_\mathsf{M}$, i.e., it holds $(\bfB_\mathsf{V})_\mathsf{M}=\bfB$.

\hspace*{-4mm}
\parbox{1\linewidth}{
\begin{infobox}{Illustration: Structure-defining matrices}
	\vspace{-3mm}
	\begin{subequations}
        \textcolor{black}{Discretised nearly constant velocity model \cite{BaLiKi:01} with position and velocity measurements has noise CMs with the special structure stemming from the first principles modelling} 
		\begin{align}
			\bfQ
			\!=\!
			\alpha_1\!\!
			\begin{bsmallmatrix}
				\frac{T_s^3}{3} &\! \frac{T_s^2}{2} \\[0.7mm] \frac{T_s^2}{2} &\! T_s
			\end{bsmallmatrix}, \ 
			\bfR
			\!=\!\!
			\begin{bsmallmatrix}
				\alpha_2 &\! \alpha_3 \\ \alpha_3 &\! \alpha_4
			\end{bsmallmatrix},
		\end{align}
		where $ T_s $ is the sampling period. The matrices \eqref{basMat} may be, in this case, defined as
        \begin{align}
			\!\!\!\!
			\bfB_Q^{(1)}\!\!&=\!\!
			\begin{bsmallmatrix}
				\frac{T_s^3}{3} &\! \frac{T_s^2}{2} \\ \frac{T_s^2}{2} &\! T_s
			\end{bsmallmatrix}, \ 
			\bfB_Q^{(2)}\!=\!\bfB_Q^{(3)}\!=\!\bfB_Q^{(4)}\!=\!\bfnul_{2\!\times\!2},
			\\
			\!\!\!\!\!\!\!\!
			\bfB_R^{(1)}\!\!&=\!\bfnul_{2\!\times\!2}, 
			\bfB_R^{(2)}\!\!=\!\!\begin{bsmallmatrix}
				1 &0 \\ 0 & 0
			\end{bsmallmatrix}, 
			\bfB_R^{(3)}\!\!=\!\!\begin{bsmallmatrix}
				0 & 1 \\ 1 & 0
			\end{bsmallmatrix}, 
			\bfB_R^{(4)}\!\!=\!\!\begin{bsmallmatrix}
				0 &0 \\ 0 & 1
			\end{bsmallmatrix}.\!\!\!\label{basisR}
		\end{align}
	\end{subequations}
\end{infobox}
}

\subsection{Residue CM Unification}

Similarly to the noise covariance $\calR_{\calE^2}$ in \eqref{noncentMom}, the residue covariance $\calR_{\tbfZ_k^2}$ also contains redundant copies of the same elements. Therefore, unique elements of the vector $\calR_{\tbfZ_k^2}$ gathered in the vector denoted as $\calR_{\tbfZ_k^2}^\rmU\in\real^{\aO(\aO+1)/ 2}$ are\linebreak selected using a unification matrix denoted as \linebreak$\Xi_{\tbfZ_k^2}\in\real^{ (\aO(\aO+1)/ 2) \times \aO^2 }$ therefore the following relationship holds
\begin{align}
	\calR_{\tbfZ_k^2}^{\rmU}=\Xi_{\tbfZ_k^2}\calR_{\tbfZ_k^2}.\label{NXiM}
\end{align}
\parbox{1\linewidth}{
\begin{infobox}{Illustration: Unification matrix}
	\vspace{-3mm}
	If $ n_{\!\zkl}\!=\!3$ and $\rank(\calO_{k})\!=\!1 $ then $ \aO^2\!=\!4 $, \linebreak$ \aO(\aO+1)/ 2\!=\!3 $ and the relation \eqref{NXiM} may be defined as
	\begin{align}
		\underbrace{\begin{bsmallmatrix}
        a_1 \\ a_2  \\ a_3
		\end{bsmallmatrix}}_{\calR_{\tbfZ_k^2}^{\rmU}}
		=\underbrace{\begin{bsmallmatrix}
				1 & 0 & 0 & 0 \\
				0 & 1 & 0 & 0 \\
				0 & 0 & 0 & 1 
		\end{bsmallmatrix}}_{\Xi_{\tbfZ_k^2}}
		\underbrace{\begin{bsmallmatrix}
			a_1 \\ a_2 \\ a_2 \\ a_3
		\end{bsmallmatrix}\!}_{\calR_{\tbfZ_k^2}}.\label{unimatrix}
	\end{align}
	Note that the unification matrix is not unique.
\end{infobox}
}

\subsection{Residue Covariance}
Substitution of \eqref{NpsiM} and \eqref{NXiM} into \eqref{noncentMom} leads to the final relation
\begin{align}
\calR_{\tbfZ_k^2}^{\rmU}&=\Xi_{\tbfZ_k^2}\calA_{k}^{\otimes^2}\calC_{k}^{\otimes^2}\Upsilon_{\calE^2}\bfalpha,\label{NmomMAllk}
\end{align}
where the vectors $\calR_{\tbfZ_k^2}^{\rmU}$ and $\bfalpha$ contain \textit{unknown} unique elements defining the residue covariance and sought noise CMs, respectively, and \textit{known} matrix $\Xi_{\tbfZ_k^2}\calA_{k}^{\otimes^2}\calC_{k}^{\otimes^2}\Upsilon_{\calE^2}$, which depends on the known system, structure-defining, and unification matrices $\bfF_k, \bfG_k, \bfE_k, \bfH_k,$ $\bfD_k, \Xi_{\tbfZ_k^2},$ $\bfB_Q^{(i)},$ and $\bfB_R^{(i)}, \forall k,\forall i$.

\section{Noise Covariance Estimation}\label{sec:mdm_identification_techniques}

The residue covariance $ \calR_{\tbfZ_k^2}^{\rmU} $ is unknown.
However, its sample-based estimate $\widehat{ \calR_{\tbfZ_k^2}^{\rmU}}= \Xi_{\tbfZ_k^2}\tbfZ_k^{\otimes^2} $ is computed from measurements and  \eqref{NmomMAllk} can be modified as
\begin{align}
	\underbrace{\Xi_{\tbfZ_k^2}\tbfZ_k^{\otimes^2}}_{\hspace{-20px}\text{Known vector}\hspace{-20px}}
	=
	\underbrace{\Xi_{\tbfZ_k^2}\calA_{k}^{\otimes^2}\calC_{k}^{\otimes^2}\Upsilon_{\calE^2}}_{\text{Known matrix}}
	\overbrace{\bfalpha^{\phantom{a}}\!\!\!}^{\hspace{-55px}\text{Sought noise cov. par. vector}\hspace{-50px}}
	+
	\underbrace{\Xi_{\tbfZ_k^2}\calA_{k}^{\otimes^2}\calC_{k}^{\otimes^2}\!}_{\text{Known matrix}}\
	\overbrace{\bfeta_k^{\phantom{a}}\!\!}^{\hspace{-20px}\text{Unknown vector}\hspace{-25px}},\hspace{-5px}\label{eq:wls2}
\end{align}
where $\bfeta_k=\calE_k^{\otimes^2}-\Upsilon_{\calE^2}\bfalpha, \forall k,$ is a \textit{zero-mean} process with the covariance and ``cross-covariance'' matrices in the vector form defined as \cite{KoDuSt:22}
\begin{subequations}\label{eq:wls3}
	\begin{align}
		\calR_{\bfeta^2}&=\mean\left[\bfeta_{k}^{\otimes^2}\right]=\mean\left[\calE_{k}^{\otimes^4}\right]-\calR_{\calE^2}^{\otimes^2},\label{eq:wls3a}\\
		\calR_{\bfeta_{k},\bfeta_{j}}&\triangleq\mean\left[\bfeta_{k}\otimes\bfeta_{j}\right]=\mean\left[\calE_{k}^{\otimes^2}\!\otimes\calE_{j}^{\otimes^2}\right]
		-\calR_{\calE^2}^{\otimes^2}.\label{eq:wls3b}
	\end{align}
\end{subequations}
with $\calR_{\calE^2}^{\otimes^2}=\left(\Upsilon_{\calE^2}\bfalpha\right)^{\otimes^2}$ due to \eqref{NpsiM}.
Writing \eqref{eq:wls2} down for all possible time instants $k=0,\ldots,\tau-\p$ leads to the following compact system of equations, which are \textit{linear} w.r.t. the sought noise covariance parameter vector $ \bfalpha $ 
\begin{align}
	\mathscr{R}_{\tbfZ^2}&=\mathscr{A}\bfalpha + \scrL\scrY,\label{NmomMAll}
\end{align}
where
\begin{subequations}
\begin{align}
	\!\!\!\!\scrR_{\tbfZ^2} \!&=\!\!
	\begin{bsmallmatrix}
			\Xi_{\tbfZ_0^2}\tbfZ_0^{\otimes^2}\\
			\Xi_{\tbfZ_{1}^2}\tbfZ_{1}^{\otimes^2}\\
			\vdots\\
			\Xi_{\tbfZ_{\tau-\p}^2}\tbfZ_{\tau-\p}^{\otimes^2}\!\!
			\\[2mm]
	\end{bsmallmatrix}\!,\
	\scrA\!=\!\!
	\begin{bsmallmatrix}
		\Xi_{\tbfZ_0^2}\calA_{0}^{\otimes^2}\calC_{0}^{\otimes^2}\\
		\Xi_{\tbfZ_{1}^2}\calA_{1}^{\otimes^2}\calC_{1}^{\otimes^2}\\
		\vdots\\
		\!\Xi_{\tbfZ_{\tau-\p}^2}\calA_{\tau-\p}^{\otimes^2}\calC_{\tau-\p}^{\otimes^2}\hspace{-2.8mm}\phantom{x}
		\\[2mm]
	\end{bsmallmatrix}\!\Upsilon_{\calE^2},
	\label{AmN}
	\\
	\!\!\scrL \!&=\!\blkdiag\!
	\begin{bsmallmatrix}
			 \!\Xi_{\tbfZ_{0}^2}\calA_{0}^{\otimes^2}\!\calC_{0}^{\otimes^2},
    &\ldots&,\
    \Xi_{\tbfZ_{\tau-\p}^2}\calA_{\tau-\p}^{\otimes^2}\calC_{\tau-\p}^{\otimes^2}\!
    \end{bsmallmatrix}\!,
    %
	\\
	\!\!\!\scrY \!&=\!\!
	\begin{bsmallmatrix}
			\bfeta_0\T, &\ \bfeta_{1}\T, &\
			\ldots &,\
			\bfeta_{\tau-\p}\T
	\end{bsmallmatrix}\T
	\!\label{Y2}.
\end{align}
\end{subequations}
and $\scrR_{\tbfZ^2}$ is a sample-based autocovariance function of $\tbfZ_k$.

The parameter vector $ \bfalpha $ can be estimated using either the \textit{ordinary} or \textit{weighted} LS methods offering a trade-off between the estimate accuracy and computational complexity. Noise CMs parameters estimation by the ordinary and weighted LS methods are discussed in Sections~\ref{sec:ols} and \ref{sec:wls}, respectively.

\subsection{Noise Identification by Ordinary LS} \label{sec:ols}
The simplest solution to the system of linear equation \eqref{NmomMAll}, assuming known full-rank matrix $\scrA$, using the ordinary LS leads to the \textit{ordinary} MDM
\begin{align}    \widehat{\bfalpha^\text{o}}=\left(\scrA\T\scrA\right)^{\!-1}\!\!\scrA\T{\scrR_{\tbfZ^2}},\label{odhadEmN}
\end{align}
where the properties of the noise $\bfeta$ are not reflected. On the other hand, the estimate is simple. Except of the snap-shot solution \eqref{odhadEmN}, we can use the \textit{recursive} LS \cite{DuKoSt:18,KoDuSt:24}.

\subsection{Noise Identification by Weighted LS}\label{sec:wls}
To improve the accuracy of the MDM and calculate the approximate CM of the estimated parameters $\bfalpha$, the weighted LS can be used instead of the ordinary LS in \eqref{odhadEmN}. The resulting method is further denoted as \textit{weighted} MDM. The weighted LS application is conditioned by the known CM of the noise component $ \scrL\bfeta $ in \eqref{NmomMAll}, which is
\begin{align}
	\mathcal{P}
	&=
	\scrL\!
	\underbrace{\!\begin{bsmallmatrix}
		(\calR_{\bfeta^2})_\mathsf{M}, & \ldots &, (\calR_{\bfeta_{\tau-\p},{\bfeta_0}})_\mathsf{M}\\
		\vdots & \ddots & \vdots\\
		(\calR_{\bfeta_{0},\bfeta_{\tau-\p}})_\mathsf{M}, & \ldots &, (\calR_{\bfeta^2})_\mathsf{M}\\
	\end{bsmallmatrix}\!}
	_{\mean\left[\scrY\scrY\T\right]}
	\!\scrL\T,\label{eq:wls5}
\end{align}
From \eqref{eq:wls3} and \eqref{eq:wls5}, it can be seen that the CM $ \mathcal{P} $ is a function of not only the sought noise covariances $ \calR_{\calE^2}$, but also of the fourth-order moments of the noises $ \mean\left[\calE_{k}^{\otimes^4}\right]$ and $ \mean\left[\calE_{k}^{\otimes^2}\otimes\calE_{j}^{\otimes^2}\right] $. These moments are, unfortunately, unknown but can be identified using the following steps:
\begin{itemize}
	\item[(1)] Compute the noise covariance parameter vector estimates by the ordinary MDM identification \eqref{odhadEmN}.
	\item[(2)] Compute fourth-order moments of the noises. If the noises are Gaussian\footnote{For the Gaussian noises, it is sufficient to estimate only the noise covariance $ \calR_{\calE^2}$ and the fourth noise moments $ \mean\big[\calE_{k}^{\otimes^4}\big]\!, \mean\big[\calE_{k}^{\otimes^2}\!\!\otimes\!\calE_{j}^{\otimes^2}\big] $\linebreak can be computed on their basis \cite{KoDuSt:22}.}, the estimated noise covariances can be used for calculation of the approximate noise moment estimates \eqref{eq:wls3}, further denoted as $\widehat{\calR_{\bfeta^2}}$, $\widehat{\calR_{\bfeta_{k},\bfeta_{j}}}$. For other distributions, the fourth moments of the noises can be estimated using the MDM as discussed in \cite{KoDuSt:22}.
	\item[(3)]  Substitute ${\calR_{\bfeta^2}}$, ${\calR_{\bfeta_{k},\bfeta_{j}}}$ by the approximate noise moments $\widehat{\calR_{\bfeta^2}}$, $\widehat{\calR_{\bfeta_{k},\bfeta_{j}}}$ in \eqref{eq:wls5}, which leads to an estimate of $\mathcal{P}$ denoted as with $\widehat{\mathcal{P}}$. 
\end{itemize}
{Then, similarly to \eqref{odhadEmN}, if $\scrA$ \color{black} and $\widehat{\mathcal{P}}$ \color{black} has the full column rank, the \textit{weighted} MDM estimate reads}
\color{black}
\begin{align}
    \widehat{\bfalpha^\text{w}}=\big(\scrA\T\widehat{\mathcal{P}}^{-1}\scrA\big)^{\!-1}\!\scrA\T\widehat{\mathcal{P}}^{-1}{\scrR_{\tbfZ^2}}.\label{eq:wls6}
\end{align}
Having the estimate of the CM $\widehat{\mathcal{P}}$, an approximate CM of the weighted MDM estimate \eqref{eq:wls6} can be computed as \color{black} \cite{MaNe:19} \color{black}
\begin{align}
	\cov\Big[\widehat{\bfalpha^\text{w}}\Big]
	\approx
	\big(\scrA\T\widehat{\mathcal{P}}^{-1}\scrA\big)^{\!-1}.\label{eq:wls7}
\end{align}
In the case of rank deficient matrix $ \widehat{\mathcal{P}} $, the weighted MDM estimate can be calculated by the LS with linear equality constraints \cite{Gi:11}. The solution, 
 obtained using the Lagrange multipliers can be written in the following compact form  \cite{MaNe:19}
\begin{align}
 \widehat{\bfalpha^\text{w}}=\Big(\scrA\T\big(\widehat{\mathcal{P}}+\scrA\scrA\T\big)^{\!\dagger}\scrA\Big)^{\!-1}\!\scrA\T\big(\widehat{\mathcal{P}}+\scrA\scrA\T\big)^{\!\dagger}{\scrR_{\tbfZ^2}},\label{eq:wls8}
\end{align}
where $ \bfA^{\dagger} $ denotes a pseudo-inverse of the matrix~$\bfA$.
An approximate (due to use of $\widehat{\mathcal{P}}$ instead of ${\mathcal{P}}$) CM of the weighted MDM estimate \eqref{eq:wls8} can be computed as 
\begin{align}
	\cov\Big[\widehat{\bfalpha^\text{w}}\Big]
	\approx
	\Big(\scrA\T\big(\widehat{\mathcal{P}}+\scrA\scrA\T\big)^{\!\dagger}\scrA\Big)^{\!-1}-\bfI_{n_\alpha}.\label{eq:wls9}
\end{align}
\color{black} If $\widehat{\mathcal{P}}\!=\!{\mathcal{P}}$ and has full rank, \eqref{eq:wls8} and \eqref{eq:wls6}, are identical \cite{MaNe:19}. \color{black}


\color{black}

\section{MDM Analysis, Extension, and Implementation}\label{sec:mdm_analysis}
\subsection{System Observability and Null-space}
The state-of-the-art noise covariance matrix identification methods are based on the assumption of system observability. In the case of the MDM version developed in \cite{DuKoSt:18, KoDuSt:22}, the observability condition is required when calculating the pseudo-inverse of the matrix $\calO_k$ to cancel the influence of the immeasurable state $\bfx_k$ in \eqref{Z}. In the proposed MDM, the state contribution is cancelled using the annihilation matrix of $\calO_k$. In this case, the matrix $\calO_k$  can be rank-deficient.

Note that the calculation of the annihilation matrix is not unique. Indeed, we can multiply \eqref{mpe_all} by a non-singular matrix $ \bfM_k\in\real^{\aO\times\aO}$, which leads to 
\begin{subequations}
	\begin{align}
			\!\!\!\bfM_k\tbfZ_k&=\bfM_k\calA_{k}\calB_{k}\begin{bsmallmatrix}-\bfU_k\\\hspace{3mm}\bfZ_k\end{bsmallmatrix}\!,\!\!\label{mpeMDMZU2}
			\\
			\!&=\bfM_k\calA_{k}\calC_{k}\begin{bsmallmatrix}\bfW_k \\ \bfV_k \end{bsmallmatrix}\!.\label{mpeMDMWV2}
		\end{align}
\end{subequations}
Then, the equation \eqref{eq:wls2} reads
\begin{multline}
	\!\!\!\!\!\!\Xi_{\tbfZ_k^2}\bfM_k^{\otimes^2}\tbfZ_k^{\otimes^2}   \!=\Xi_{\tbfZ_k^2}\bfM_k^{\otimes^2}\calA_{k}^{\otimes^2}\calC_{k}^{\otimes^2}\Upsilon_{\calE^2}
	\bfalpha
	\\+
	\Xi_{\tbfZ_k^2}\bfM_k^{\otimes^2}\calA_{k}^{\otimes^2}\calC_{k}^{\otimes^2}
	\bfeta_k,\!\!\!\!\label{mver}
\end{multline}
Eqn. \eqref{mver} can be interpreted as \eqref{eq:wls2} multiplied by a non-singular matrix $\bfM_k^{\otimes^2}$ as shown in the following relation for left-hand-sides of \eqref{mver} and \eqref{eq:wls2}
\begin{align}
 %
 \underbrace{\Xi_{\tbfZ_k^2}\bfM_k^{\otimes^2}\tbfZ_k^{\otimes^2}}_{\text{Eq. \eqref{mver}}}
	=
	\underbrace{\ \Xi_{\tbfZ_k^2}\bfM_k^{\otimes^2}\Psi_{\tbfZ_k^2}\  }_{\text{non-singular matrix}}\hspace{0.5mm}
    \underbrace{\Xi_{\tbfZ_k^2}\tbfZ_k^{\otimes^2}}_{\text{Eq. \eqref{eq:wls2}}}.
\end{align}
Thus, the choice of the matrix $ \bfM_k$ affects the estimation using the ordinary MDM \eqref{odhadEmN} but the effect will be negligible\footnote{If the matrix $\mathcal{P}$ were not replaced by an estimate in \eqref{eq:wls6}, the choice of matrix $ \bfM_k$ does not affect the CM estimate.} when using the weighted MDM \eqref{eq:wls6}.
The matrix $\Psi_{\tbfZ_k^2}\in\real^{\aO^2 \times (\aO(\aO+1)/ 2)}$ is the replication matrix \cite{KoDuSt:22} which conveniently duplicates unique elements and can be understood as the counterpart of the unification matrix $\Xi_{\tbfZ_k^2}$~\eqref{NXiM} i.e., 
$\tbfZ_k^{\otimes^2}=\Psi_{\tbfZ_k^2}\Xi_{\tbfZ_k^2}\tbfZ_k^{\otimes^2}$. 

\hspace*{-4mm}
\parbox{1\linewidth}{
\begin{infobox}{Illustration: Replication matrix}
    \vspace*{-3mm}
	If $ \aO^2\!=\!4 $, $ \aO(\aO+1)/ 2\!=\!3 $, and \eqref{unimatrix} holds then
	\begin{align}
		\underbrace{\begin{bsmallmatrix}
			a_1 \\ a_2 \\ a_2 \\ a_3
		\end{bsmallmatrix}\!}_{\tbfZ_k^{\otimes^2}}
		=\underbrace{\begin{bsmallmatrix}
				1 & 0 & 0 \\
				0 & 1 & 0 \\
                0 & 1 & 0 \\
				0 & 0 & 1 
		\end{bsmallmatrix}}_{\Psi_{\tbfZ_k^2}}
        \overbrace{
        \underbrace{\begin{bsmallmatrix}
				1 & 0 & 0 & 0 \\
				0 & 1 & 0 & 0 \\
				0 & 0 & 0 & 1 
		\end{bsmallmatrix}}_{\Xi_{\tbfZ_k^2}}
        \underbrace{\begin{bsmallmatrix}
            a_1 \\ a_2 \\ a_2  \\ a_3
		\end{bsmallmatrix}}_{\tbfZ_k^{\otimes^2}}}^{\begin{bsmallmatrix}
            a_1 & a_2 & a_3
		\end{bsmallmatrix}\T}.
	\end{align}
 It should be highlighted that ${\Psi_{\tbfZ_k^2}\Xi_{\tbfZ_k^2}}\neq{\bfI_4}$.
\end{infobox}}
\subsection{Residue Definition for Unknown Input}
The residue calculation $\tbfZ_k$ \eqref{mpe_all} requires availability of the input $\bfu_k, \forall k$. However, this assumption can be relaxed by the following calculation of the residue
\begin{align}
	\tbfZ_k
	&= \anni\!\left(\begin{bmatrix} \calO_{k},& \bfGamma_k\calG_k \end{bmatrix}\right) \bfZ_k,
	\\
	&=\anni\!\left(\begin{bmatrix} \calO_{k},& \bfGamma_k\calG_k \end{bmatrix}\right)\begin{bmatrix}\bfGamma_k, & \bfI_{n_{\zkl}}\end{bmatrix}\calC_{k}\begin{bsmallmatrix}\bfW_k \\ \bfV_k \end{bsmallmatrix},
\end{align}
where the influence of the input vector $\bfU_k$ is cancelled by the multiplication of the augmented measurement vector by an annihilation matrix of the state and input transformation matrix, i.e., $\begin{bmatrix} \calO_{k},& \bfGamma_k\calG_k \end{bmatrix}$. This leads to a modification of $\calA_{k}\in\real^{\aOG\times (\p n_x+n_{\zkl})}$, with \linebreak$\aOG\!=\!n_{\zkl}\!-\!\rank(\begin{bmatrix} \calO_{k},& \bfGamma_k\calG_k \end{bmatrix})$, as
\begin{align}
    \calA_{k}=\anni\!\left(\begin{bmatrix} \calO_{k},& \bfGamma_k\calG_k \end{bmatrix}\right)\begin{bmatrix}\bfGamma_k, & \bfI_{n_{\zkl}}\end{bmatrix}\label{eq:AunknownU}
\end{align}
used in \eqref{mpeMDMWV} and the following equations. Then, the noise CM elements can be estimated analogously using the ordinary or weighted MDM. 

Note that the noise covariance estimation under an unknown input sequence has been treated in \cite{KoSuArChTiBiWe:23} for time-invariant and detectable models only.



\subsection{Noise CM Elements Identifiability}
{In \cite{KoDuSt:22, KoDuSt:21}, the number of identifiable parameters was analysed for all correlation methods under the assumptions of time-invariant and observable model, and a relation between the state and measurement dimensions $n_x, n_z$ and the number of identifiable elements of the noise CMs $\bfQ, \bfR$ was derived.  This analysis is also valid for the \textit{proposed} ordinary and weighted MDM versions assuming $\calO_k$ of full-rank.}

{Unfortunately, abandoning these assumptions complicates the derivation of such a relation as}
\begin{itemize}
    \item {For the LTV models, the derivation is hardly manageable in practice as the identifiability depends on the possibly time-varying measurement dimension $n_{z_k}$.}
    \item {For unobservable systems, the number of identifiable parameters is also a function of possibly time-varying observability matrix rank $\rank(\calO_k)$.}
\end{itemize}
{In summary, the analytical relation determining the number of identifiable parameters of the noise CMs $\bfQ, \bfR$ as a function of $n_x, n_{z_k}, \rank(\calO_k)$ has yet to be established.}

{Nevertheless, the number of the noise CMs identifiable parameters, \textcolor{black}{i.e., the number of the unbiased parameter estimates}, can easily be inferred \textit{numerically} by calculating the rank of $\mathscr{A}$ \eqref{NmomMAll}, which depends on the \textit{known} model matrices $\bfF_k, \bfG_k, \bfE_k, \bfH_k, \bfD_k, \forall k,$ only.}

{It is important to note that the number of identifiable elements is also affected by the availability or unavailability of the control input. An unknown control input may \textit{reduce} the number of identifiable parameters. In a case of equal input and state noise shaping matrices $\bfG_k=\bfE_k, \forall k$, the annihilation matrix used in \eqref{eq:AunknownU} cancels not only the influence of the input but also of the state noise. As a consequence, the state noise CM $\bfQ$ is not, in this case, identifiable at all}.

\subsection{Estimate Properties}\label{sec:estProp}
{Like other implicit correlation methods that do not explicitly calculate the state estimate, the ordinary MDM \eqref{odhadEmN} provides unbiased and consistent estimates. The proof is similar to the one presented in \cite{KoDuSt:22}. The proposed generalised version of the MDM offers unbiased and consistent estimates of the noise CMs $\bfQ, \bfR$ elements provided that the matrix $\mathscr{A}$ \eqref{NmomMAll} has full column rank as discussed above. The weighted MDM \eqref{eq:wls6}, however, produces biased estimates in general.} The reason is the calculation of the CM $\mathcal{P}$ \eqref{eq:wls5} from the same data, which are later used for the estimate calculation in \eqref{eq:wls6}. The estimate bias goes to zero with an increasing number of data.

\subsection{User-Defined Parameter Selection}
{The introduced generalised MDM requires specification of one user-defined parameter $\p$ in the augmented measurement vector creation \eqref{Z}. The parameter $\p$ affects not only the existence of the annihilation matrix $am(\calO_k)$ in \eqref{eq:am}, but also the number of equations in \eqref{odhadEmN}.}

{Existence of the annihilation matrix is ensured if the dimension of the augmented measurement vector $\bfZ_k$, which depends on $\p$, exceeds the rank of the observability matrix $\rank(\calO_k)$ in \eqref{eq:matrices}. It means, $\p$ should be selected such that the condition\footnote{The observability matrix rank is always less than (or equal to, for observable systems) $n_x$.} $n_{z,k,L}>\rank(\calO_k)$ is satisfied $\forall k$. This condition can thus be used for specification of the minimum value of $\p$. The user may, however, select $\p$ to be greater than this minimum value, resulting in an increased number of equations for the ordinary (i.e., \textit{unweighted}) LS solution \eqref{odhadEmN}. However, a higher number of equations increases  computational complexity and may not improve the noise CMs estimate in \eqref{odhadEmN} as discussed in \cite{DuKoStBl:20} for the MDM.}

\subsection{Extension of MDM Features}
The generalised MDM (based on the annihilation matrix) can straightforwardly use the capabilities and extensions of already published versions of the MDM, which include
\begin{itemize}
    \item Noise mean and higher-order moments identification \cite{KoDuSt:22},
    \item Parameters of the time/mutually correlated noises \cite{KoDuSt:18b},
    \item Parameters of the noise probability density functions \cite{KoDuSt:18a},
    \item Noise Gaussianity assessment for state space model \cite{DuKoStBl:20},
    \item Identification of measurement noise covariance only, i.e., sensor calibration with unknown state dynamic \cite{KoDuStDa:23},
    \item {User-defined parameter selection \cite{DuKoStBl:20}}.
\end{itemize}

\section{Available MDM source codes}\label{sec:source}
\MATLAB implementation of the MDM versions developed in this paper
can be found at:\\\scalebox{0.95}{\url{https://github.com/IDM-UWB/NoiseIdentification_MDM}}.

Based on the provided model matrices, measurement, and alternatively input vectors (defined in Section 2.1) for all time instants, the source code computes all the necessary matrices and vectors used by the MDM in \eqref{eq:wls2} i.e., $ \Xi_{\tbfZ_k^2}\tbfZ_k^{\otimes^2} $, $ \Xi_{\tbfZ_k^2}\calA_{k}^{\otimes^2}\calC_{k}^{\otimes^2}\Upsilon_{\calE^2}$, and $\Xi_{\tbfZ_k^2}\calA_{k}^{\otimes^2}\calC_{k}^{\otimes^2}$. 
In addition, the code provides the covariances and ``cross-covariances'' of $ \bfeta_k$ for weighted LS in \eqref{eq:wls5} as functions of the (estimated) noise covariances $ \calR_{\calE^2} $ \eqref{eq:wls3}  i.e., $\calR_{\bfeta_{k+j},{\bfeta_k}}=\bff_j(\calR_{\calE^2})$, for $ j=0,1,\hdots ,\p $, under assumption of Gaussian densities.

Besides the MDM version for the LTV model, a version for the linear time-invariant (LTI) model was also published, which significantly speeds up the calculation of the necessary matrices.








\section{Numerical Illustrations}\label{sec:numerical_illustration}
The MDM is illustrated using the following three examples
\begin{itemize}
    \addtolength{\itemindent}{-3mm}
    \item Unobservable LTI model of a clock ensemble (Sec. \ref{eq:exam1}),
    \item Unobservable LTV model with unknown input (Sec. \ref{eq:exam2}),
    \item Observable LTV model (Sec. \ref{eq:exam3}).
\end{itemize}
Whereas the first two examples validate the proposed generalised MDM for unobservable models, which has not been solved in the literature before, the third example allows its comparison with other approaches using observable models. The performance of the noise covariance identification is assessed using $10^4$ Monte-Carlo (MC) simulations with $\tau=10^3$ measurement samples per MC simulation. All considered models \eqref{model} have the Gaussian initial state being Gaussian with $\mean\left[x_0\right]=\mathbf{1}_{n_x\times1}$ and $\var\left[x_0\right]=\bfI_{n_x}$, the input is $u_k=\sin(k/\tau)$ with dimension $ n_{u_k}=1$. The state and measurement noises are independent and generated according to the Gaussian density.

\subsection{Unobservable LTI System: Ensemble of Clocks}\label{eq:exam1}
{The illustration starts with the model used in calculating the stable time scale based on the combination of outputs of an ensemble of atomic clocks \cite{Galleani2008,Galleani2010}. Each clock state is characterised by the time and frequency deviations, where time deviation is experimentally observed as a Brownian motion-type noise}. As a consequence, the deviations of an ensemble of three clocks can be modelled using \eqref{eqSt} and \eqref{eqMe} with
\begin{subequations}
\begin{align}
	\!\!\!\bfF_k&=\left(\bfI_3 \otimes
    \begin{bsmallmatrix}
		1 & T_s \\[0.7mm] 0 & 1
	\end{bsmallmatrix}\right)\!,\
    \bfG_k=\bfnul_{6\times1},\
    \bfE_k=\bfI_6,
    \\
	\!\!\!\!\!\!\!\!\!\bfH_k&=\begin{bsmallmatrix}
	    1 & 0 & -1 & 0 & \hspace{2.8mm}0 & 0 \\ 1 & 0 & \hspace{2.8mm}0 & 0 & -1 & 0 
	\end{bsmallmatrix}\!,\
	\bfD_k=\bfI_2,\ T_s\hspace{-0.2mm}=\hspace{-0.2mm}10,
\end{align}
\begin{align}
    \!\!\!\!\!\!\bfQ&=10^{-19}\!\cdot\!\!
    \left(
    \begin{matrix}
    \hspace{-15mm}
    \begin{bsmallmatrix} 6&0& 0\\0 & 20 & 0\\0 & 0 &7 \end{bsmallmatrix}\!\!\otimes\!\!
    \begin{bsmallmatrix}
		\frac{T_s^3}{3} &\!\! \frac{T_s^2}{2} \\[0.7mm] \frac{T_s^2}{2} &\!\! T_s
	\end{bsmallmatrix}
    \\[-3mm] \\ \hspace*{13mm}+
    \!\begin{bsmallmatrix} 0.05&0&0\\0 & 0.3 &0\\0 &0 &0.04 \end{bsmallmatrix}\!\!\otimes\!\!
    \begin{bsmallmatrix}
		T_s &\!\! 0 \\ 0 &\!\! 0
	\end{bsmallmatrix}\!\!
    \end{matrix}
    \right)\!\!,\!\!\!\label{eq:ex1Q}
    \\
    \!\!\!\!\!\!\bfR&=10^{-19}\!\cdot\!\!\begin{bsmallmatrix}
	    80&\!\! 0\\0 &\!\! 100
	\end{bsmallmatrix},
    \ n_x\hspace{-0.2mm}=\hspace{-0.2mm}n_w\hspace{-0.2mm}=\hspace{-0.2mm}6,\ n_{z_k}\!=\hspace{-0.2mm}n_v\hspace{-0.3mm}=\hspace{-0.2mm}2.\!\!\label{eq:ex1R}
\end{align}
\end{subequations}
\textcolor{black}{The state and measurement noise CMs $\bfQ$ and $\bfR$ in \eqref{eq:ex1Q} and \eqref{eq:ex1R}, respectively, have known structures derived from the first-principles modelling \cite{Galleani2008}. They can be described using user-selected\footnote{{Considered definition of the structure-defining matrices is a user choice and thus, is not unique.}} structure-defining matrices $\bfB_Q^{(i)}, \bfB_R^{(j)}, i=1, \ldots,6, j=1,2,$ and the vector of weights $\bfalpha$ fulfilling the form \eqref{basMat} to be \textit{estimated}. The considered structure-defining matrices are}
\begin{subequations}
\begin{align}
    \!\!\!\!\bfB_Q^{(1)}&\!\!=\!\!
    \begin{bsmallmatrix} 1&0&0\\0 &0 & 0\\0 & 0 & 0 \end{bsmallmatrix}
    \!\!\otimes\!\!
    \begin{bsmallmatrix}
		\frac{T_s^3}{3} &\!\! \frac{T_s^2}{2} \\[0.7mm] \frac{T_s^2}{2} &\!\! T_s
	\end{bsmallmatrix},\
    \bfB_Q^{(2)}\!\!=\!\!
    \begin{bsmallmatrix} 1& 0& 0\\0 & 0 &0\\0 & 0 & 0 \end{bsmallmatrix}
    \!\!\otimes\!\!
    \begin{bsmallmatrix}
		T_s &\!\! 0 \\ 0 &\!\! 0
	\end{bsmallmatrix},\!\!
    \\
    \!\!\!\!\bfB_Q^{(3)}&\!\!=\!\!
    \begin{bsmallmatrix} 0&0& 0\\0 &1 & 0\\0 &0 & 0 \end{bsmallmatrix} \!\!\otimes\!\!
    \begin{bsmallmatrix}
		\frac{T_s^3}{3} &\!\! \frac{T_s^2}{2} \\[0.7mm] \frac{T_s^2}{2} &\!\! T_s
	\end{bsmallmatrix},\
    \bfB_Q^{(4)}\!\!=\!\!
    \begin{bsmallmatrix} 0&0& 0\\0 & 1 & 0\\0 & 0 & 0 \end{bsmallmatrix} \!\!\otimes\!\!
    \begin{bsmallmatrix}
		T_s &\!\! 0 \\ 0 &\!\! 0
	\end{bsmallmatrix},\!\!
    \\
    \!\!\!\!\bfB_Q^{(5)}&\!\!=\!\!
    \begin{bsmallmatrix} 0& 0& 0\\0 &0 &0\\0 & 0 & 1 \end{bsmallmatrix} \!\!\otimes\!\!
    \begin{bsmallmatrix}
		\frac{T_s^3}{3} &\!\! \frac{T_s^2}{2} \\[0.7mm] \frac{T_s^2}{2} &\!\! T_s
	\end{bsmallmatrix},\
    \bfB_Q^{(6)}\!\!=\!\!
    \begin{bsmallmatrix} 0& 0&0\\0 &0 &0\\0 & 0 & 1 \end{bsmallmatrix} \!\!\otimes\!\!
    \begin{bsmallmatrix}
		T_s &\!\! 0 \\ 0 &\!\! 0
	\end{bsmallmatrix},\!\!
    \\
    \!\!\!\!\!\!\!\!\!\bfB_R^{(1)}&\!=\bfB_R^{(2)}\!=\bfB_R^{(3)}\!=\bfB_R^{(4)}\!=\bfB_R^{(5)}\!=\bfB_R^{(6)}\!=\bfnul_{2\times2},
    \\
    \!\!\!\!\!\!\!\!\!\bfB_Q^{(7)}\!&=\bfB_Q^{(8)}\!=\bfnul_{6\times6},\ \ 
    \bfB_R^{(7)}\!=\!\begin{bsmallmatrix}
        1& 0 \\0& 0
    \end{bsmallmatrix},\
    \bfB_R^{(8)}\!=\!\begin{bsmallmatrix}
        0& 0\\0& 1
    \end{bsmallmatrix},\!\!
    \label{ex1:basisQR}
\end{align}
\end{subequations}
{and the weights are}
\begin{align}
    \bfalpha\!=\!10^{-19}\times\!\begin{bmatrix} 6,&\!\!\!\! 0.05,&\!\!\!\! 20,&\!\!\!\! 0.3,&\!\!\!\! 7,&\!\!\!\! 0.04,&\!\!\!\! 80,&\!\!\!\! 100 \end{bmatrix}\T.
\end{align}

%

\begin{figure}[h]
	\hspace{-3mm}
	\includegraphics[scale=0.7]{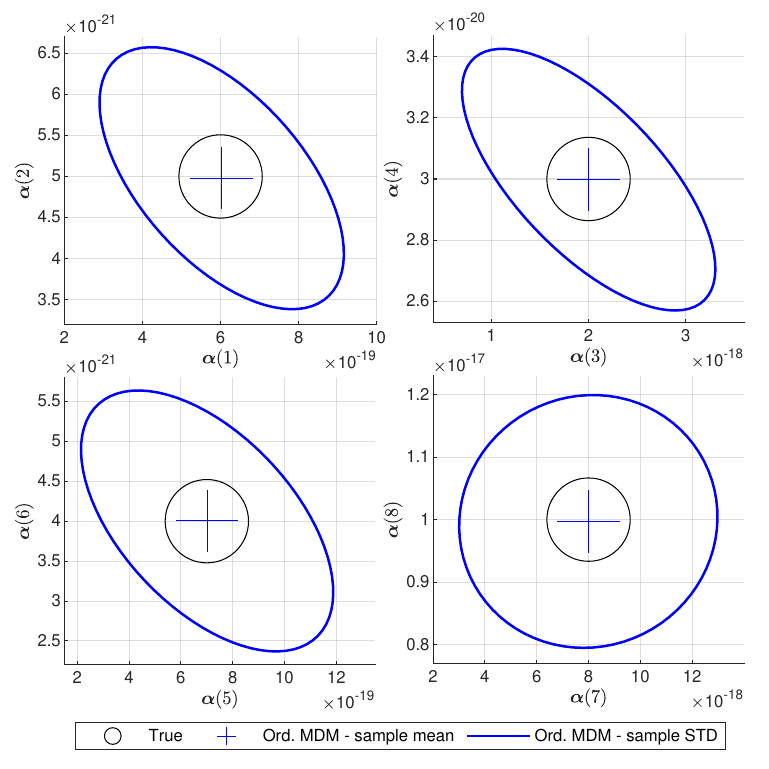}
	\vspace*{-4mm}
	\caption{Ordinary MDM identification (unobservable, LTI).}
	\label{UnObsUnCon}
	\vspace*{-1mm}
\end{figure}

The sample mean (S. mean) and the diagonal of the sample covariance matrix (S. cov) of ordinary MDM estimates for $L\!=\!10$ over the MC simulations are shown in Table~\ref{tableUnObs} and depicted in Fig.~\ref{UnObsUnCon}, where STD stands for the standard deviation. Table \ref{tableUnObs} and Fig.~\ref{UnObsUnCon} confirm that the ordinary MDM estimate is unbiased. 
\begin{table}[h]
	\vspace{-0mm}
	\hspace{-1mm}
	\begin{tabular}{cccccccc}
		& \!\!\!\!\! & \hspace{-7px} $\widehat{\bfalpha^\text{o}}(1)$ & $\widehat{\bfalpha^\text{o}}(2)$ & $\widehat{\bfalpha^\text{o}}(3)$ & $\widehat{\bfalpha^\text{o}}(4)$ 
		\\[-11px]
		\\ \cmidrule{1-1}\cmidrule{3-6} 
		\\[-10px]
		\hspace{-10px}True\hspace{-10px} &\!\!\!\!\!\! & \!\!\!\!6$\cdot10^{-19}$ & 5$\cdot10^{-21}$ & 2$\cdot10^{-18}$ & 3$\cdot10^{-20}$\!\!\!\! \\
		\hspace{-10px} S.\! mean\hspace{-10px}  &\!\!\!\!\!\! & \!\!\!\!6.028$\cdot10^{-19}$ & 4.979$\cdot10^{-21}$ & 2.001$\cdot10^{-18}$ & 2.998$\cdot10^{-20}$\!\!\!\! \\
        \hspace{-4px} S.\! cov \hspace{-0px} &\!\!\!\!\!\! & \!\!\!\!9.785$\cdot10^{-38}$ & 2.543$\cdot10^{-42}$ & 1.701$\cdot10^{-36}$ & 1.830$\cdot10^{-41}$\!\!\!\!
        \\ \cmidrule{1-1}\cmidrule{3-6} 
    	\\[-5px]
		& \!\!\!\!\! & \hspace{-7px} $\widehat{\bfalpha^\text{o}}(5)$ & $\widehat{\bfalpha^\text{o}}(6)$ & $\widehat{\bfalpha^\text{o}}(7)$ & $\widehat{\bfalpha^\text{o}}(8)$
		\\[-11px]
		\\ \cmidrule{1-1}\cmidrule{3-6} 
		\\[-10px]
		\hspace{-10px}True\hspace{-10px} &\!\!\!\!\!\! & \!\!\!\!7$\cdot10^{-19}$ & 4$\cdot10^{-21}$ & 8$\cdot10^{-18}$ & 1$\cdot10^{-17}$\!\!\!\! \\
		\hspace{-10px} S.\! mean \hspace{-10px} &\!\!\!\!\!\! & \!\!\!\!7.011$\cdot10^{-19}$ & 4.003$\cdot10^{-21}$ & 7.992$\cdot10^{-18}$ & 0.997$\cdot10^{-17}$\!\!\!\! \\
        \hspace{-4px} S.\! cov \hspace{-0px} &\!\!\!\!\!\! & \!\!\!\!2.369$\cdot10^{-37}$ & 2.670$\cdot10^{-42}$ & 2.485$\cdot10^{-35}$ & \!\!4.091$\cdot10^{-36}$\!\!\!\!
        \\ \cmidrule{1-1}\cmidrule{3-6} 
	\end{tabular}
	\caption{Ordinary MDM identification for unobservable clock model.}
	\label{tableUnObs}
	\vspace*{-0mm}
\end{table}
\subsection{Unobservable LTV System with Unknown Input}\label{eq:exam2}
In the second example, we focus on the illustration of the generalised MDM for an unobservable system with \textit{unknown} input. The model \eqref{eqSt}, \eqref{eqMe} has the following parameters
\begin{subequations}
\begin{align}
	\!\!\!\!\!\bfF_k&\hspace{-0.8mm}=\hspace{-1.4mm}\begin{bsmallmatrix}
	    1 & 2 & 1\\ 0 & -1.01 & 2\\ 0 & 0 & 1
	\end{bsmallmatrix},\
    \bfG_k\hspace{-0.8mm}=\hspace{-1.4mm}\begin{bsmallmatrix}
	    0\\ \sin\left(\!\frac{10k}{\tau}\!\right)\!\\ 1
	\end{bsmallmatrix},\
    \bfE_k\hspace{-0.8mm}=\hspace{-1.4mm}\begin{bsmallmatrix}
	    -3&2&0\\ 2&2&2\\ 5&0&1
	\end{bsmallmatrix},\!\!\!\!\!
    \\
	\!\!\!\!\!\!\!\!\!\bfH_k&\hspace{-0.8mm}=\hspace{-1.4mm}\begin{bsmallmatrix}
	    0 &1 & 0\\ 0 &0 & 2\\ 0 &1 & 1
	\end{bsmallmatrix},\
	\bfD_k\hspace{-0.8mm}=\hspace{-1.4mm}\begin{bsmallmatrix}
	    1&1& \hspace{2.1mm}0\\ 0&2& \hspace{2.1mm}1\\ 1&0&-1
	\end{bsmallmatrix},\
    \bfQ\hspace{-0.8mm}=\hspace{-1.4mm}\begin{bsmallmatrix}
	    1 &1 &0\\ 1 &2 & 1\\ 0 & 1 &2
	\end{bsmallmatrix},\\
    \!\!\bfR&\hspace{-0.8mm}=\hspace{-1.4mm}\begin{bsmallmatrix}
	    2 & 0 & 1\\ 0 & 4 & 1\\ 1 & 1 & 2
	\end{bsmallmatrix},\
	n_x=n_w=n_{z_k}=n_v=3,
\end{align}
\end{subequations}
and the structure-defining matrices can be defined as
\begin{align}
    \!\!\!\!\!\bfB_Q^{(1)}&\!=\!\!\begin{bsmallmatrix}
        1&0& 0\\0& 1& 0\\0& 0&\!\! 1
    \end{bsmallmatrix},\hspace{0.5mm}
    \bfB_Q^{(2)}\!=\!\!\begin{bsmallmatrix}
        0& 0& 0\\0& 1& 0\\0& 0& 1
    \end{bsmallmatrix},\hspace{0.5mm}
    \bfB_Q^{(3)}\!=\!\!\begin{bsmallmatrix}
        \hspace{2mm}0& -1& \hspace{2mm}0\\-1& \hspace{2mm}0 &-1\\ \hspace{2mm}0& -1& \hspace{2mm}0
    \end{bsmallmatrix},\!\!\!\!
    \\
    \!\!\!\!\!\!\bfB_R^{(4)}&\!=\!\!\begin{bsmallmatrix}
        1&0&0\\0& 0&0\\0& 0& 1
    \end{bsmallmatrix},\hspace{0.5mm}
    \bfB_R^{(5)}\!=\!\!\begin{bsmallmatrix}
        0&0&0\\0&2& 0\\0&0& 0
    \end{bsmallmatrix},\hspace{0.5mm}
    \bfB_R^{(6)}\!=\!\!\begin{bsmallmatrix}
        0&0&1\\0& 0 &1\\ 1& 1& 0
    \end{bsmallmatrix},\!\!\!\!
    \\
    \!\!\!\!\!\!\bfB_Q^{(4)}&\!=\bfB_Q^{(5)}\!=\bfB_Q^{(6)}\!=\bfB_R^{(1)}\!=\bfB_R^{(2)}\!=\bfB_R^{(3)}\!=\bfnul_{3\times3}.
    \label{ex2:basisQR}
\end{align}
Due to \eqref{basMat}, the vector of parameters is
$\bfalpha\!=\!\begin{bmatrix} 1, 1, -1, 2, 2, 1 \end{bmatrix}\T$.

%

\begin{figure}[h]
	\hspace{-2mm}
	\includegraphics[scale=0.61]{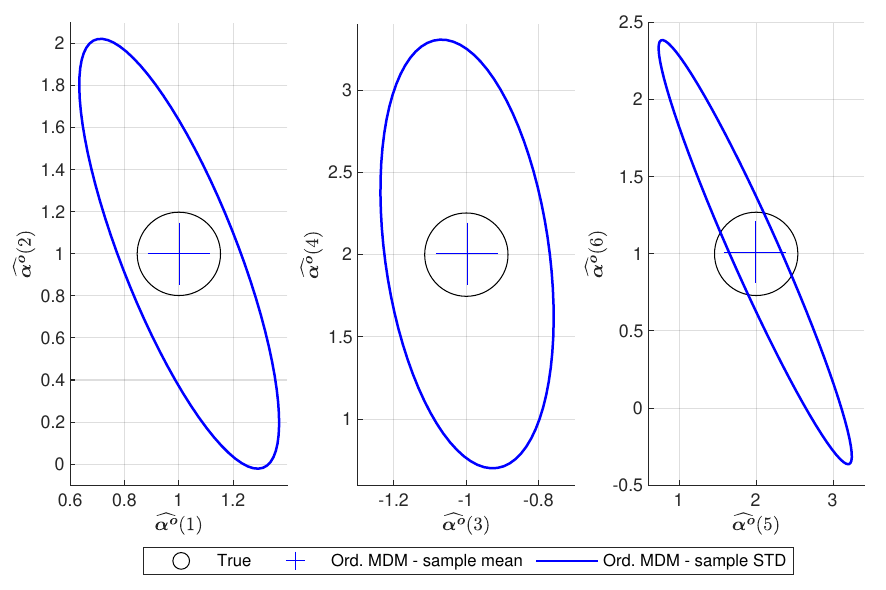}
	\vspace*{-3mm}
	\caption{Ordinary MDM identification (unobservable, LTV, unknown input).}
	\label{ex3:UnObsUnCon}
	\vspace*{-1mm}
\end{figure}

The S. mean and S. cov of ordinary MDM estimates for $L=2$ over the MC simulations are shown in Table \ref{ex3:tableUnObs} and depicted in Fig.~\ref{ex3:UnObsUnCon}. 
\begin{table}[h]
	\vspace{-0mm}
	\hspace{-0mm}
	\begin{tabular}{ccccccccc}
		& \!\!\!\!\! & \hspace{-7px} $\widehat{\bfalpha^\text{o}}(1)$ & $\widehat{\bfalpha^\text{o}}(2)$ & $\widehat{\bfalpha^\text{o}}(3)$ & $\widehat{\bfalpha^\text{o}}(4)$ & $\widehat{\bfalpha^\text{o}}(5)$ & $\widehat{\bfalpha^\text{o}}(6)$
		\\[-11px]
		\\ \cmidrule{1-1}\cmidrule{3-8} 
		\\[-10px]
		\hspace{-4px}True\hspace{-4px} &\!\!\!\!\!\! & 1 & 1 & -1 & 2 & 2 & 1 \\
		\hspace{-7px} S.\! mean  &\!\!\!\!\!\! & 1.001 & 1.001 & -0.998 & 2.004 & 1.987 & 1.011\\
        \hspace{-14px} S.\! cov \hspace{-14px} &\!\!\!\!\!\! & 0.135 & 1.045 & 0.057 & 1.695 & 1.574 & 1.893
        \\ \cmidrule{1-1}\cmidrule{3-8} 
	\end{tabular}
	\caption{Ordinary MDM identification for unobservable system and unknown input.}
	\label{ex3:tableUnObs}
	\vspace*{-0mm}
\end{table}

\subsection{Observable LTV System: Correlation Methods Comparison}\label{eq:exam3}
In the third example, we consider the observable LTV system adopted from \cite{KoDuSt:22}\footnote{The sign in the measurement equation is incorrectly written in \cite{KoDuSt:22}. Note, however, that the impact of the sign on the results is negligible.} to confront the proposed generalisations with previous versions of the MDM and the method proposed by \cite{Be:74}. The considered model \eqref{eqSt} and \eqref{eqMe} with 
\begin{subequations}
\begin{align}
	F_k&=0.8-0.1\sin(7\pi k / \tau),\ \ G_k=E_k=D_k=1,\\
	H_k&=1+0.99\sin(100\pi k / \tau),\ \ Q=2,\ \ R=1,\\
	n_x&=n_w=n_{z_k}=n_v=1,
\end{align}
\end{subequations}
and the  structure-defining matrices are 
\begin{align}
    B_Q^{(1)}=1, \ B_Q^{(2)}=0,\ \ \ \
    B_R^{(1)}=0, \ B_R^{(2)}=1
    \label{ex1:basisR}
\end{align}
and due to \eqref{basMat} is
$Q\!=\!\bfalpha(1)\!=\!2$ and $R\!=\!\bfalpha(2)\!=\!1$, and $L=2$.

\begin{figure}[h]
	\hspace{-2mm}
	\includegraphics[scale=0.58]{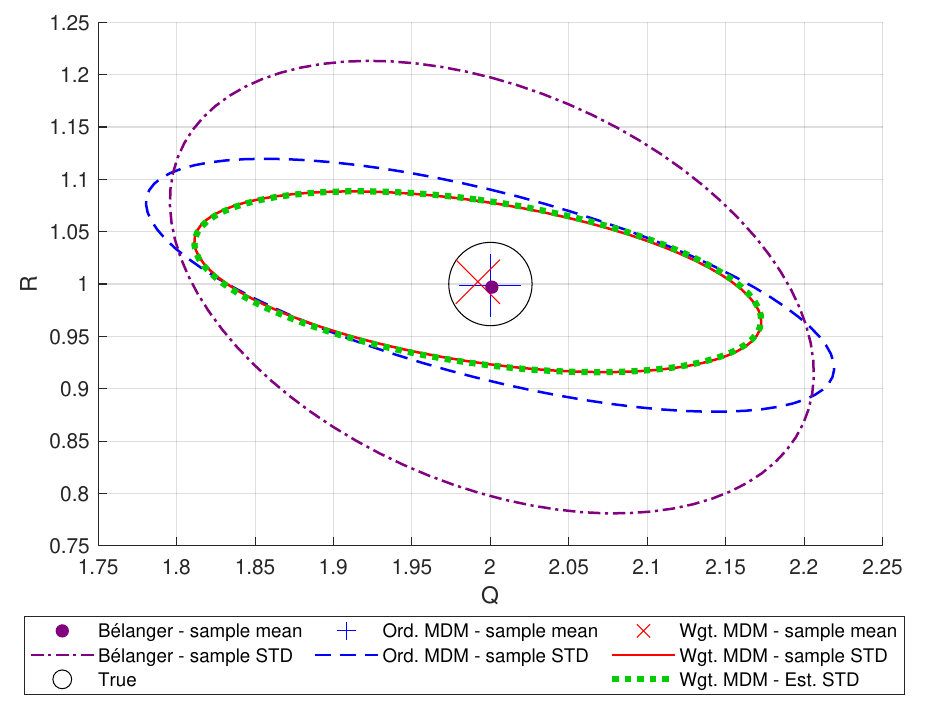}
	\vspace*{-4mm}
	\caption{Bélanger's method and (ordinary and weighted) MDM identification.}
	\label{tacFig}
	\vspace*{-0mm}
\end{figure}

The estimate S. mean and S. cov of the ordinary MDM, weighted MDM, and the correlation method 
given in \cite{Be:74} over the MC simulations are shown in Table \ref{tableTAC} and depicted in Fig.~\ref{tacFig}. In addition, an average computation time and an average of the covariance matrix diagonal of $ \big(\scrA\T\widehat{\mathcal{P}}^{\!-1}\scrA\big)^{\!-1}$ \eqref{eq:wls7} over the MC simulations (Est. cov) are shown in Table \ref{tableTAC}.  
Estimate covariance matrix calculation, i.e., Est. cov, has not been discussed for the ordinary MDM and Bélanger's method \cite{Be:74}; thus, it is also not considered in this section.
%
%
%
%
%
%
%
%
%
%
\begin{table}[h]
	\vspace{-1mm}
	\hspace{-1mm}
    \scalebox{1.02}{  
	\begin{tabular}{cccccccccccccccc}
		& &\!\!\!\!\!& \multicolumn{2}{c}{Bélanger} & \!\!\!\!\!\! & \multicolumn{2}{c}{\hspace{-7px}Ordinary MDM\hspace{-17px}} \!\!\!\!\!\! & \!\!\!\!\!\! & \multicolumn{3}{c}{\hspace{-4px}Weighted MDM}\\
		& \hspace{-4px}True\hspace{-4px} & \!\!\!\!\!\!\!\!\!
        & \hspace{-9px} S.\! mean & \hspace{-17px} S.\! cov \hspace{-17px} & \!\!\!\!\!\! 
		& \hspace{-9px} S.\! mean & \hspace{-17px} S.\! cov \hspace{-17px} & \!\!\!\!\!\! 
		& \hspace{-9px} S.\! mean & \hspace{-17px} S.\! cov \hspace{-17px} & \hspace{-4px} Est.\! cov\hspace{-6px}
		\\[-11px]
		\\ \cmidrule{1-2}\cmidrule{4-5} \cmidrule{7-8} \cmidrule{10-12}  
		\\[-10px]
		\!\!\!$\bfalpha(1)$\!\!\!\! & \!2\!   &\!\!\!\!\! & \!\!\!\!2.001  &  \hspace{-5px}0.042\hspace{-5px} &\!\!\!\!\! & \!\!\!\!2.000  &  \hspace{-5px}0.048\hspace{-5px} &\!\!\!\!\! & \!\!\!\!\!\!\!1.992\!\!\!\!  &  \hspace{-3px}0.033\hspace{-3px} & 0.033\!\!\!\!
        \\
		\!\!\!$\bfalpha(2)$\!\!\!\! & \!1\!   &\!\!\!\!\!\! & \!\!\!\!0.997  &  \hspace{-5px}0.047\hspace{-5px} &\!\!\!\!\!\! & \!\!\!\!0.999  &  \hspace{-5px}0.015\hspace{-5px} &\!\!\!\!\!\! & \!\!\!\!\!\!\!1.002\!\!\!\!  &  \hspace{-5px}0.007\hspace{-5px} & 0.007\!\!\!\!
		\\[-11px]
		\\ \cmidrule{1-2}\cmidrule{4-5} \cmidrule{7-8} \cmidrule{10-12}  
		\\[-10px]
		\!\!\!\!\text{Time}\!\!\!\! & - &\!\!\!\!\!\!\!\! &   \multicolumn{2}{c}{56$\cdot10^{-3}$s} &\!\!\!\!\!\!\!\! &   \multicolumn{2}{c}{\hspace{-5px}0.735$\cdot10^{-3}$s\hspace{-5px}}   &\!\!\!\!\!\! &    \multicolumn{3}{c}{\hspace{-5px}1.759s\hspace{-5px}} \\
	\end{tabular}
    }
	\caption{Performance of ordinary and weighted MDM identification.}
	\label{tableTAC}
	\vspace*{-3mm}
\end{table}

Fig.~\ref{tacFig} shows that the weighted MDM provides an estimate with significantly lower covariance than the ordinary MDM estimate and also than Bélanger's method. Moreover, the weighted MDM provides a realistic assessment of the estimate uncertainty, which means that the calculated estimate covariance, or the STD in this case, \eqref{eq:wls7} is the same as the sample covariance calculated from the MC estimates. {However, the weighted MDM provides estimates with a small \textit{bias} as discussed in Section \ref{sec:estProp}.}

Note that the weighted MDM \eqref{eq:wls6} provides almost identical S. mean and S. cov as the weighted MDM \cite{KoDuSt:22} based on pseudo-inverse of the matrix $\calO_k$.


\subsection{Assessment of Results}
The numerical illustrations confirm that the ordinary MDM provides unbiased estimates for any linear model. The weighted MDM provides a realistic evaluation of the estimate covariance matrix. The MDM and any other correlation methods have a similar computational complexity \cite{DuStKoHa:17}. The weighted MDM requires additional computational overhead to evaluate the weighting matrix. Comparison of the MDM, developed in \cite{DuKoSt:18}, with other noise covariance estimation methods can be found in e.g., \cite{KoDuSt:24} and references therein. 

%
\section{Concluding Remarks}\label{sec:conclusion}
The paper focused on the noise identification of stochastic linear time-varying state-space models. A general problem was considered, where the dimension of the measurements and inputs may vary in time. Compared to the state-of-the-art methods, the proposed generalised measurement difference method does not rely on system observability or detectability and input availability assumptions. {The proposed MDM generalisation led to two algorithms with differing weighting matrices in the underlying least-squares method, allowing for a trade-off between estimate accuracy and computational complexity}. The performance of the proposed techniques was analysed and discussed using three models with diverse properties. Implementation of all the proposed techniques in \MATLAB is publicly available. {Like other state-of-the-art noise identification correlation methods, the proposed MDM generalisation estimates the same number of the noise CMs elements for observable systems. For unobservable systems, the number of identifiable noise CMs elements decreases. While it is possible to determine the number of identifiable elements for unobservable systems numerically, the analytical relationship that infers these elements from the dimensions of state and measurement vectors, and the properties of the state and measurement matrices remains an unresolved question.}

\bibliography{literatura}

\begin{thebibliography}{10}
\providecommand{\url}[1]{#1}
\csname url@samestyle\endcsname
\providecommand{\newblock}{\relax}
\providecommand{\bibinfo}[2]{#2}
\providecommand{\BIBentrySTDinterwordspacing}{\spaceskip=0pt\relax}
\providecommand{\BIBentryALTinterwordstretchfactor}{4}
\providecommand{\BIBentryALTinterwordspacing}{\spaceskip=\fontdimen2\font plus
\BIBentryALTinterwordstretchfactor\fontdimen3\font minus
  \fontdimen4\font\relax}
\providecommand{\BIBforeignlanguage}[2]{{%
\expandafter\ifx\csname l@#1\endcsname\relax
\typeout{** WARNING: IEEEtran.bst: No hyphenation pattern has been}%
\typeout{** loaded for the language `#1'. Using the pattern for}%
\typeout{** the default language instead.}%
\else
\language=\csname l@#1\endcsname
\fi
#2}}
\providecommand{\BIBdecl}{\relax}
\BIBdecl

\bibitem{GaSaClGo:22}
R.~Gao, S.~Särkkä, R.~Claveria-Vega, and S.~Godsill, ``Autonomous tracking
  and state estimation with generalized group lasso,'' \emph{IEEE Transactions
  on Cybernetics}, vol.~52, no.~11, pp. 12\,056--12\,070, 2022.

\bibitem{Meh:72}
R.~K. Mehra, ``Approaches to adaptive filtering,'' \emph{IEEE Transactions on
  Automatic Control}, vol.~17, no.~10, pp. 693--698, 1972.

\bibitem{DuStKoHa:17}
J.~Dun\'{i}k, O.~Straka, O.~Kost, and J.~Havl\'{i}k, ``Noise covariance
  matrices in state-space models: {A} survey and comparison - part {I},''
  \emph{International Journal of Adaptive Control and Signal Processing},
  vol.~31, no.~11, pp. 1505--1543, 2017.

\bibitem{SaNu:09}
S.~S\"{a}rkk\"{a} and A.~Nummenmaa, ``Recursive noise adaptive {K}alman
  filtering by variational {B}ayesian approximations,'' \emph{IEEE Transactions
  on Automatic Control}, vol.~54, no.~3, pp. 596--600, 2009.

\bibitem{HuZyShCh:21}
Y.~Huang, Y.~Zhang, P.~Shi, and J.~Chambers, ``Variational adaptive {K}alman
  filter with {G}aussian-{I}nverse-{W}ishart mixture distribution,'' \emph{IEEE
  Transactions on Automatic Control}, vol.~66, no.~4, pp. 1786--1793, 2021.

\bibitem{MoKi:93}
A.~R. Moghaddamjoo and R.~L. Kirlin, \emph{Approximate Kalman Filtering}, ser.
  Series in Approximations and Decompositions.\hskip 1em plus 0.5em minus
  0.4em\relax World scientific, 1993, vol.~2, ch. Robust Adaptive Kalman
  Filtering, pp. 65--85, (Ed. Chen, G.).

\bibitem{SoHaIlAbBi:14}
A.~Solonen, J.~Hakkarainen, A.~Ilin, M.~Abbas, and A.~Bibov, ``Estimating model
  error covariance matrix parameters in extended {K}alman filtering,''
  \emph{Nonlinear Processes in Geophysics}, vol.~21, no.~5, pp. 919--927, 2014.

\bibitem{ShSt:00}
R.~H. Shumway and D.~S. Stoffer, \emph{Time Series Analysis and its
  Applications}.\hskip 1em plus 0.5em minus 0.4em\relax Springer-Verlag, 2000.

\bibitem{ScWiNi:11}
T.~B. Sch{\"o}n, A.~Wills, and B.~Ninness, ``System identification of nonlinear
  state-space models,'' \emph{Automatica}, vol.~47, no.~1, pp. 39--49, 2011.

\bibitem{LuLePu:21}
M.~M. Lucini, P.~J. van Leeuwen, and M.~Pulido, ``Model error estimation using
  the expectation maximization algorithm and a particle flow filter,''
  \emph{SIAM/ASA Journal on Uncertainty Quantification}, vol.~9, no.~2, pp.
  681--707, 2021.

\bibitem{CoKl:24}
N.~Cohen and I.~Klein, ``{A-KIT}: Adaptive {K}alman-informed transformer,''
  \emph{arXiv: 2401.09987}, 2024.

\bibitem{Be:74}
P.~R. B{\'e}langer, ``Estimation of noise covariance matrices for a linear
  time-varying stochastic process,'' \emph{Automatica}, vol.~10, no.~3, pp.
  267--275, 1974.

\bibitem{OdeRaRa:06}
B.~J. Odelson, M.~R. Rajamani, and J.~B. Rawlings, ``A new autocovariance
  least-squares method for estimating noise covariances,'' \emph{Automatica},
  vol.~42, no.~2, pp. 303--308, 2006.

\bibitem{KoDuSt:22}
O.~Kost, J.~Duník, and O.~Straka, ``Measurement difference method: A universal
  tool for noise identification,'' \emph{IEEE Transactions on Automatic
  Control}, vol.~68, no.~3, pp. 1792--1799, 2023.

\bibitem{Galleani2008}
L.~Galleani, ``{A tutorial on the two-state model of the atomic clock noise},''
  \emph{Metrologia}, vol.~45, no.~6, pp. 175--182, 2008.

\bibitem{Galleani2010}
L.~Galleani and P.~Tavella, ``{Time and the Kalman Filter: Applications of
  optimal estimation to atomic timing},'' \emph{IEEE Control Systems Magazine},
  vol.~30, no.~2, pp. 44--65, 2010.

\bibitem{BaLiKi:01}
Y.~Bar-Shalom, X.~R. Li, and T.~Kirubarajan, \emph{Estimation with Applications
  to Tracking and Navigation: Theory Algorithms and Software}.\hskip 1em plus
  0.5em minus 0.4em\relax John Wiley \& Sons, 2001.

\bibitem{LiDeCh:19}
J.~Li, F.~Deng, and J.~Chen, ``A fast distributed variational {B}ayesian
  filtering for multisensor ltv system with non-{G}aussian noise,'' \emph{IEEE
  Transactions on Cybernetics}, vol.~49, no.~7, pp. 2431--2443, 2019.

\bibitem{DuKoSt:18}
J.~Dun\'{i}k, O.~Kost, and O.~Straka, ``Design of measurement difference
  autocovariance method for estimation of process and measurement noise
  covariances,'' \emph{Automatica}, vol.~90, pp. 16--24, 2018.

\bibitem{KoDuSt:24}
O.~Kost, J.~Dun\'{i}k, and O.~Straka, ``Noise covariances identification by
  {MDM}: Weighting, recursion, and implementation,'' in \emph{Accepted for the
  20th IFAC Symposium on System Identification {\nobreak (SYSID)}}, Boston
  (MA), USA, 2024.

\bibitem{MaNe:19}
J.~R. Magnus and H.~Neudecker, \emph{Matrix Differential Calculus with
  Applications in Statistics and Econometrics, Revised Edition}, 3rd~ed.\hskip
  1em plus 0.5em minus 0.4em\relax John Wiley \& Sons, 2019.

\bibitem{Gi:11}
B.~P. Gibbs, \emph{Advanced {K}alman Filtering, Least-Squares and
  Modelling}.\hskip 1em plus 0.5em minus 0.4em\relax John Wiley \& Sons, 2011.

\bibitem{KoSuArChTiBiWe:23}
H.~Kong, S.~Sukkarieh, T.~J. Arnold, T.~Chen, B.~Mu, and W.~X. Zheng,
  ``Identifiability analysis of noise covariances for {LTI} stochastic systems
  with unknown inputs,'' \emph{IEEE Transactions on Automatic Control},
  vol.~68, no.~7, pp. 4459--4466, 2023.

\bibitem{KoDuSt:21}
O.~Kost, J.~Dun\'{i}k, and O.~Straka, ``Identifiability of unique elements of
  noise covariances in state-space models,'' in \emph{Proceedings of 19th IFAC
  Symposium on System Identification {\nobreak (SYSID)}}, Padova, Italy, 2021.

\bibitem{DuKoStBl:20}
J.~Dun\'{i}k, O.~Kost, O.~Straka, and E.~Blasch, ``Covariance estimation and
  gaussianity assessment for state and measurement noise,'' \emph{Journal of
  Guidance, Control, and Dynamics}, vol.~43, no.~1, pp. 132--139, 2020.

\bibitem{KoDuSt:18b}
O.~Kost, J.~Dun\'{i}k, and O.~Straka, ``Correlated noise characteristics
  estimation for linear time-varying systems,'' in \emph{57th Conference on
  Decision and Control}, Miami, FL, USA, Dec 2018.

\bibitem{KoDuSt:18a}
------, ``Noise moment and parameter estimation of state-space model,''
  \emph{In Proceedings of 18th IFAC Symposium on System Identification
  {\nobreak (SYSID)}}, vol.~51, no.~15, pp. 891--896, 2018.

\bibitem{KoDuStDa:23}
O.~Kost, J.~Duník, O.~Straka, and O.~Daniel, ``Identification of {GNSS}
  measurement error: From time to elevation dependency,'' \emph{IEEE
  Transactions on Aerospace and Electronic Systems}, vol.~59, no.~6, pp.
  8931--8943, 2023.

\end{thebibliography}
\bibliographystyle{IEEEtran}

\end{document}